\documentclass[%
 reprint,
 amsmath,amssymb,
 aps,
]{revtex4-2}

\usepackage{graphicx}
\usepackage{dcolumn}
\usepackage{bm}
\usepackage{hyperref}

\begin{document}

\preprint{APS/123-QED}

\title{Tuning the Topology of a Two-Dimensional Catenated DNA Network}

 
\author{Indresh Yadav$^1$}
\author{Dana Al Sulaiman$^2$}
\author{Patrick S. Doyle$^1$}
 \altaffiliation{pdoyle@mit.edu}
\affiliation{
 $^1$ Department of Chemical Engineering, Massachusetts Institute of Technology, Cambridge, Massachusetts 02139, United States \\
 $^2$ Division of Physical Science and Engineering, King Abdullah University of Science and Technology, Thuwal 23955-6900, Saudi Arabia 
}%


\begin{abstract}
Molecular topology of polymers plays a key role in determining their physical properties. We studied herein the topological effects on the static and dynamic properties of a 2D catenated network of DNA rings called a kinetoplast. Restriction enzymes, that cleave DNA at sequence-specific sites, are used to selectively cut and remove rings from the network and hence tune the molecular topology while maintaining overall structural integrity. We find that topology has minimal effects over the spatial extension of the 2D network, however it significantly affects the relaxation behavior. The shape fluctuations of the network are governed by two distinct characteristic time scales attributed to the thermal fluctuations and confinement of the network. The relationship between the time constant of thermal relaxation and the amplitude of anisotropy fluctuations yields a universal scaling. Interestingly, this scaling is independent of the detailed arrangements of rings and/or perforation within the catenated networks. This study provides a route to tune  the elastic properties of 2D catenated DNA networks by modifying the underlying topology in a rational and highly controllable manner. 
\end{abstract}

\maketitle

The topology of polymer molecules plays an important role in determining their static (e.g., size and shape) and dynamic (e.g., relaxation and diffusion) properties \cite{Wulstein2019,Soh2019}. Tuning the topology of molecules is thus imperative in controlling the physiochemical properties tailored for desired applications  \cite{frank1981topological}. 
Recently, a class of mechanically interlinked polymers molecules called polycatenanes have received considerable attention because of their unique rheological, mechanical, and thermal properties \cite{Hart2021a,liu2022polycatenanes,lang2014swelling}. Catenated ring networks, or Olympic gels \cite{Raphael1997}, have been developed with muscle-like properties \cite{Hu2020} and reconfigurable topologies \cite{Kim2013,Krajina2018}. 
 To understand the physics of catenated ring networks and hence engineer them for a desired application, a robust model system is required wherein topology can be tuned and studied at the single molecule level. Nature provides such a model system in the form of giant 2D catenated ring networks called kinetoplasts \cite{Shapiro1995,Chen1995}.
 


Kinetoplast DNA (kDNA) from trypanosomatid \textit{Crithidia fasciculata} is a natural Olympic gel wherein approximately 5000 minicircles  ($\sim 2.5 $ kbp) and 25 maxicircles  ($\sim 40 $ kbp) are topologically interlocked in a quasi-2D plane \cite{Chen1995}. The catenation valency of minicircles is approximately three\cite{Chen1995}. Maxicircles are also linked to each other, however the catenation valency is unknown. While the network of maxicircles and minicircles are interlocked with each other, each network can be sustained independently \cite{Shapiro1993a}. The nucleotide sequence of minicircles and maxicircles is different, and this property can be exploited to selectively digest maxicircles or minicircles in a controllable manner. Moreover, there are two different classes of minicircles, namely the major class (accounting for about $\sim 90 \%$ of the network) and the minor class (accounting for about $\sim 10 \%$ of the network), where each class is homogeneous in its base pair sequence but different from the other class \cite{Birkenmeyer1985}. There is currently no understanding of how these various classes of rings affect the kDNA conformation, polymer dynamics or mechanical properties.

Recently our group has studied various aspects of kDNA and proposed it as a model system for 2D catenated polymers \cite{Klotz2020,Soh2020b,Soh2020a,Soh2021,Yadav2021}. For example, it has been shown that in response to constriction or electric field stretching, kDNA behave as an elastic sheet \cite{Klotz2020,Soh2020b}. Its properties as a 2D polyelectrolyte have also been studied in response to the degree of confinement and ionic strength \cite{Soh2020a,Soh2021}. The unsatisfactory match of scaling laws based on a generalized Flory approach for a 2D polymer indicates that the physics of catenated polymers is yet to be fully realized. Furthermore, our group has demonstrated that the characteristic of polymer-and-salt-induced phase transition of kDNA is quite different than that of linear DNA \cite{Yadav2021}. 

\begin{figure*}[t]
\centering
\includegraphics[width=0.9\linewidth]{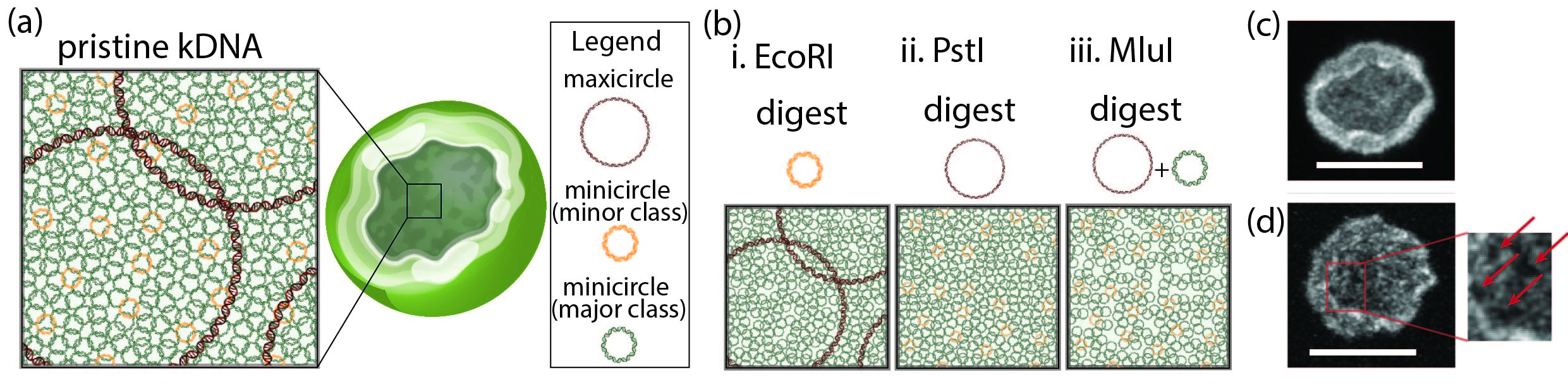}
\caption{(a) Schematic diagram of a kDNA, a catenated 2D network (Olympic gel) made of thousands of connected rings of circular DNA. (b) With the precise action of restriction enzymes, either ring class or a combination of classes of rings can be removed from the network, hence selectively tuning the topology. Super resolution confocal microscopy image (c) showing the pristine kDNA shape as a wrinkled bowl with positive Gaussian curvature, (d) a representative molecule after digestion by the \textit{MluI} enzyme for 10 minutes, where digestion creates observable perforations within the network. Scale bar is $5~\mu\mathrm{m}$.}
    \label{fig:fig1}
\end{figure*}

In this Letter, we study the equilibrium size, shape and relaxation spectrum of  kDNA as a function of its molecular topology. A number of restriction enzymes were used to selectively remove the rings from  the catenated network. Time-controlled digestion was also performed to randomly remove rings from the network to varying degrees in a controllable manner. Using this approach, we could selectively tune the topology of the kDNA while maintaining its overall molecular integrity, and thus generated a library of molecules with different topologies. Our results show that while topology has minimal effects on the kDNA size and shape, it significantly affects the relaxation behavior owing to the mechanical strength of the kDNA network. We envision that this strategy can provide opportunities to selectively tune the physical properties of catenated DNA networks and shed light over the much debated physics of 2D polymers. The results also provide rich information and guide the design of future 2D materials with specific properties. 


The heterogeneity in base pair sequencing of different classes of rings and the precise action of restriction enzymes have been harnessed to tune the topology of kDNAs. A summary of the experimental methodology is presented in Figure 1. The restriction enzyme \textit{EcoRI} has been used to cut the minor class of minicircles \cite{Birkenmeyer1985,Hoeijmakers1982}, \textit{PstI} to cut the maxicircles \cite{Carpenter1995}, and a combination of \textit{EcoRI} and \textit{PstI} to cut both the minor class of minicircles and maxicircles. Next, we used \textit{MluI} to digest the major class of minicircles as well as maxicircles \cite{Birkenmeyer1985} where fractional digestion of the network was controlled by digestion time. Single molecule experiments were conducted in straight microfluidic channels with a $2~\mu\mathrm{m}$ height, $40~\mu\mathrm{m}$ width and a 1 cm length. Our prior work \cite{Klotz2020,Yadav2021} showed that this channel height orients kDNAs for ease of imaging and only moderately confines it. See Supplemental Material (SM) for experimental details\cite{SM2022}. 

Figure 2 presents a montage of fluorescence microscopy images of kDNAs. The ionic strength of the solution is 32.3 mM \cite{Soh2020a} which corresponds to a Debye length of 1.69 nm compared to the 2 nm bare width of dsDNA. Figure 2(a) shows representative images of control (pristine) kDNA and kDNA remodeled using various enzymes. The image displays a single frame (frame number 1137, chosen arbitrarily). Each row displays the typical variability in  shape and size of kDNA molecules.  All the molecules have a nearly elliptical shape with a clearly identified outline. The brightness of the outline (or edge) is attributed to the dense fibril network of rings at the periphery of kDNAs \cite{Chen1995}. This observation is consistent with previous work from our group \cite{Klotz2020,Soh2021}. These molecules have a topologically flat conformation with positive Gaussian curvature \cite{Yadav2021,Polson2021}. Interestingly, we observe qualitatively that there is no major difference in the shape and size of the molecules after removal of the minor class of minicircles, the maxicircles, or a combination of both.  

The influence of molecular topology was further quantified by measuring the radius of gyration ($R_{\mathrm{g}}$) from the 2D projection of the fluorescence intensity of each kDNA, which maps kDNA's extension in a plane \cite{Soh2020b,Yadav2021}. Here, $R_{\mathrm{g}}$ was determined as the square root of the trace of the radius of gyration tensor \cite{Hsieh2007}. Violin plots, representing the distribution of $R_{\mathrm{g}}$, are presented in Figure 2(c). Each violin plot maps the probability distribution of $R_{\mathrm{g}}$, where the width of a plot corresponds to the frequency of the data points in each region. The central solid line represents the median, while dotted lines represent the ends of first and third quartiles. Statistical testing comparing the different distributions with the \textit{control} gives $p > 0.05$, which implies that modifying kDNA topology by removing the minor class of minicircles and/or the maxicircles does not significantly influence the molecule's size. This non-trivial result implies, interestingly, that among the components of the kDNA network, it is the major class of the minicircles that plays the most important role in dictating the overall molecular size. This is also evidence that the catenated polymers offer great variability in their topology but very robust geometrical structure.
\begin{figure}[h]
\centering
\includegraphics[width=0.7\linewidth]{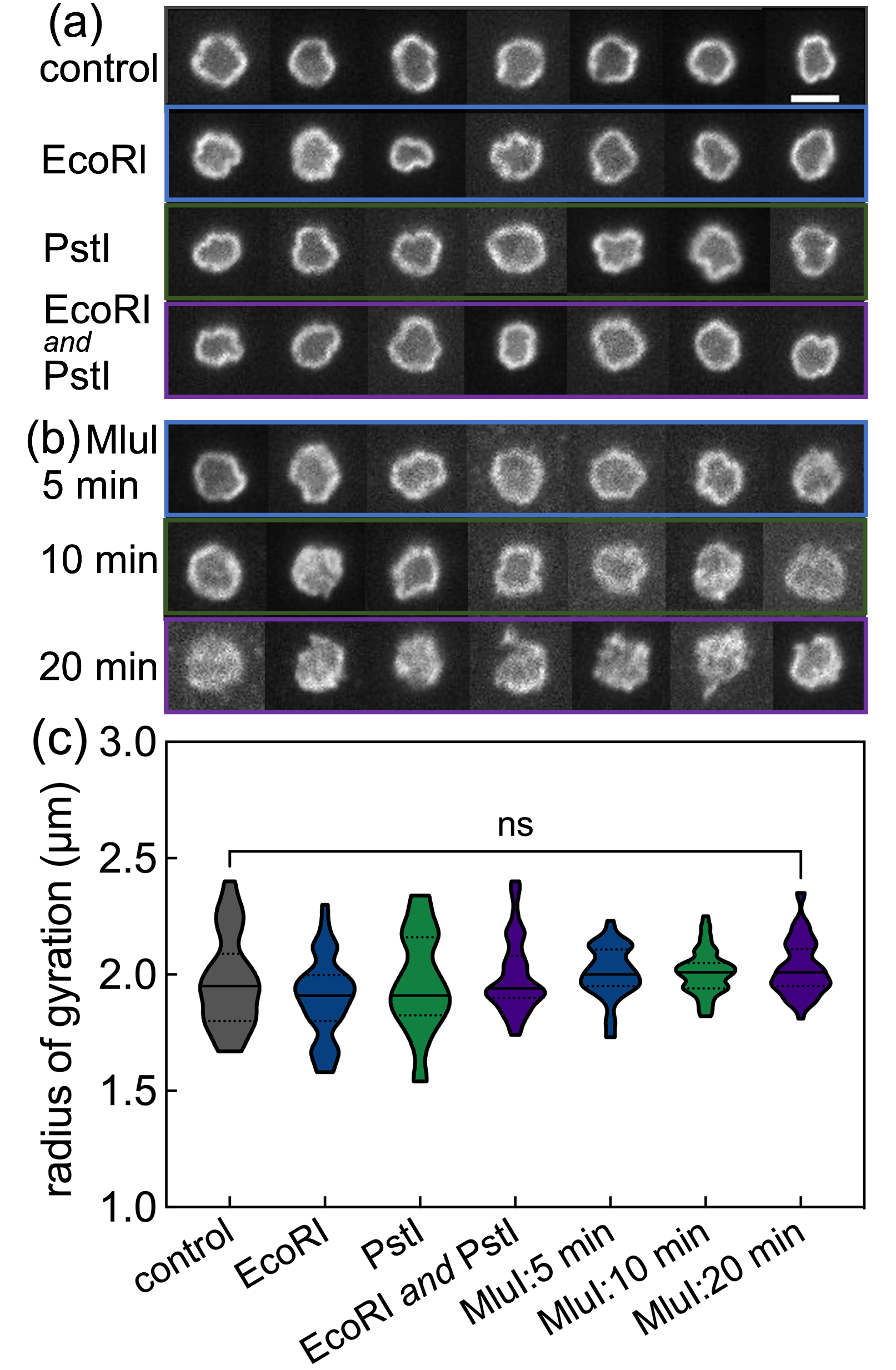}
    \caption{(a) and (b) Fluorescence microscopy images of kDNAs corresponding to control and kDNAs remodelled with restriction enzymes. (c) Violin plots of the size distribution of kDNAs corresponding to control  and kDNAs remodelled with restriction enzymes. Scale bar is $5~\mu\mathrm{m}$.}
    \label{fig:fig2}
\end{figure}

To investigate this result further, we chose to tune the amount of the major class of minicircles. While there is no enzyme which can independently digest this type of ring, we used \textit{MluI} which digests both the major class of minicircles and the maxicircles. Representative images of molecules corresponding to different degrees of digestion, which is a function of digestion time, is shown in Figure 2(b). Molecules corresponding to a digestion time of $\ge$ 10 minutes show a change in shape compared to the control. For a reaction time of 20 minutes, below which molecular integrity is intact, significant visual deviation in shape of the molecules can be seen. The $R_{\mathrm{g}}$ distributions are presented in Figure 2(c). Again, with regards to the molecular size of kDNA, statistical testing shows no significant difference in the $R_{\mathrm{g}}$ distribution.  Another interesting featured is that we did not observe the crumpling of molecules even with the maximum digestion we could achieve while still maintaining the molecular integrity. 

Irrespective of their topology, kDNA molecules undergo small shape fluctuations (Movie S1 and S2) and for a given environmental condition fluctuation in shape is stationary. 
Here, thermal fluctuation is measured in terms of variation in anisotropy as a function of time for individual molecules. The shape of a kDNA can be described as wrinkled bowl with an elliptical cross-section \cite{Klotz2020}. Principle eigenvalues of the gyration tensor calculated from the projected fluorescence intensity give the length of the minor and major axes of the ellipse. The anisotropy then is defined as the ratio of the minor axis and the major axis. The anisotropy measured in this way provides information about symmetry where a value of $1$ represents circular conformation and a value of $0$ a straight line. Hence, studying the temporal behaviour of anisotropy not only provides the degree of thermal fluctuation but also the robustness/flexibility of the shape of the network manifold. The distribution of anisotropy for molecules in different samples are presented in Figure S3. Though there is significant molecule-to-molecule variation in anisotropy within a sample, the average value for different samples remains almost same ($p>0.05$). This also implies that the shape of the molecule does not change drastically by tuning its topology. The amplitude of fluctuation relative to its mean value, however, is significantly affected by tuning the topology. Variation in anisotropy as a function of time for quintessential molecules for the \textit{control} and a \textit{MluI}-digested molecule are presented in Figure 3(a). The amplitude of anisotropy fluctuation for a given molecule is quantified in terms of the standard deviation of anisotropy ($\sigma$) over a given time series. It is clear that $\sigma$ value increases after removing a fraction of rings from the network. The ensemble averaged standard deviation ($\sigma$) of the anisotropy calculated using equation S1 for different samples is presented in Table S1.

\begin{figure}[h]
\centering
\includegraphics[width=0.7\linewidth]{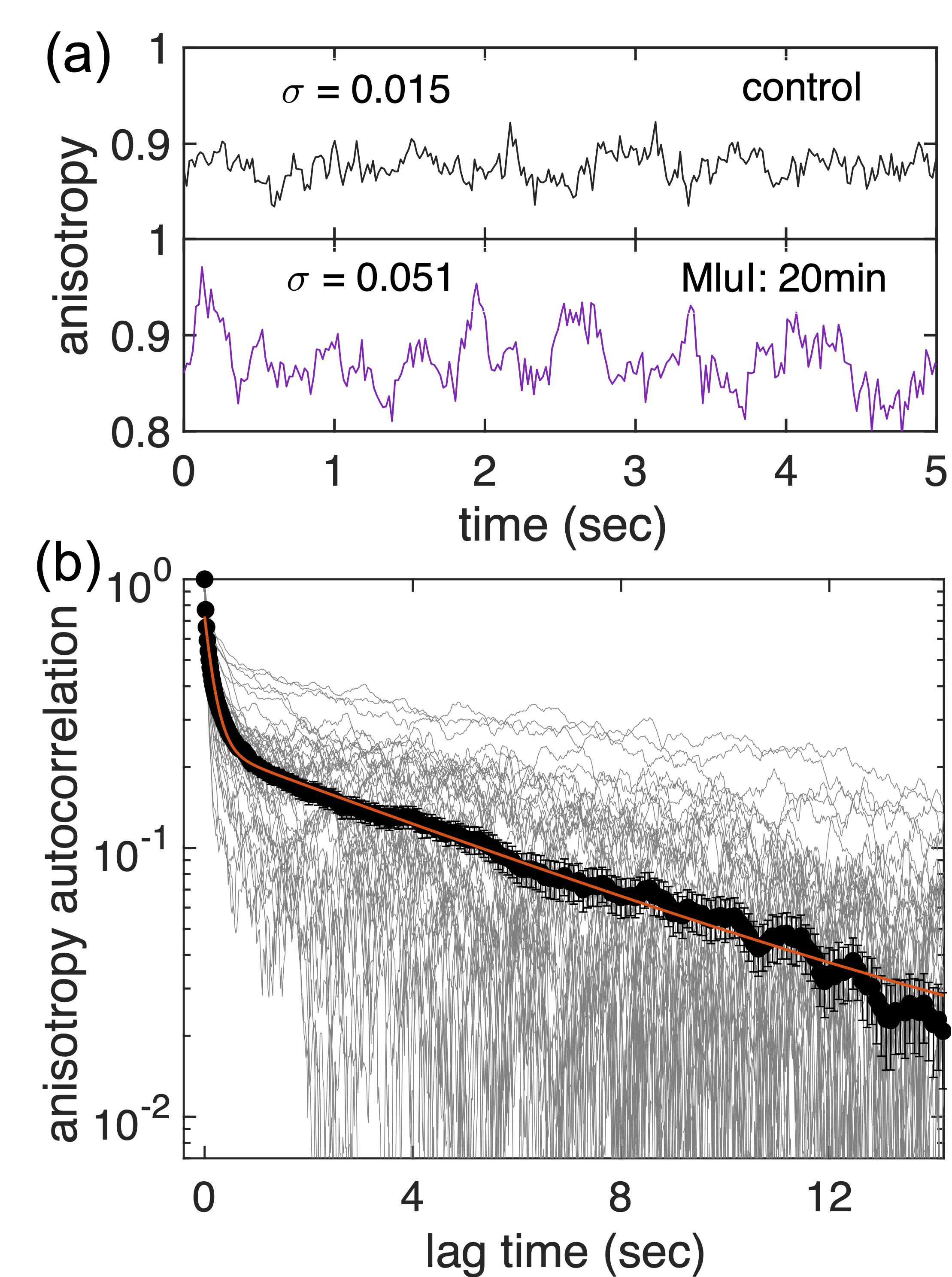}
    \caption{(a) Upper panel shows the anisotropy fluctuation for control (pristine) kDNA and the bottom panel for a molecule digested with MluI for 20 min. Though the mean value of anisotropy remains almost the same ($\sim0.9$), the standard deviation ($\sigma$) varies for different topologies. (b) Semilogarithmic plot of temporal autocorrelation function of anisotropy for the control molecules (N=49) and their ensemble average (black).}
    \label{fig:fig3}
\end{figure}
To study the dynamic behavior of the kDNA network we calculated the autocorrelation function \textit{$C(\tau)$} of the anisotropy \textit{A} measured over \textit{T} camera frames  
\begin{equation}
C(\tau)=\frac{1}{\sigma^{2} T} \sum_{t=1}^{T-\tau}(A(t)-\bar{A})(A(t+\tau)-\bar{A})
\end{equation}
where $\sigma^{2}$ is the anisotropy variance and $\bar{A}$ is the average value of anisotropy over a given time series. An ensemble of autocorrelation functions and their average for control molecules is presented in Figure 3(b). Qualitatively the autocorrelation for each molecule has a fast decay at short lag time followed by a tail at longer lag times. Accordingly, the ensemble average autocorrelation is fitted with the sum of two exponentials [SM, Equation S2]. The fast time scale ($\tau_1$) was found to be $0.177 \pm 0.018 \mathrm{~s}$ (SD), which is of the order of the relaxation timescale for the much smaller $\lambda$-DNA \cite{Hsieh2007,Reisner2005}. The characteristic timescale of the longer time tail ($\tau_2$) was found to be $ 5.90 \pm 0.81\mathrm{~s}$ (SD).   

\begin{figure*}[t]
\centering
\includegraphics[width=1\linewidth]{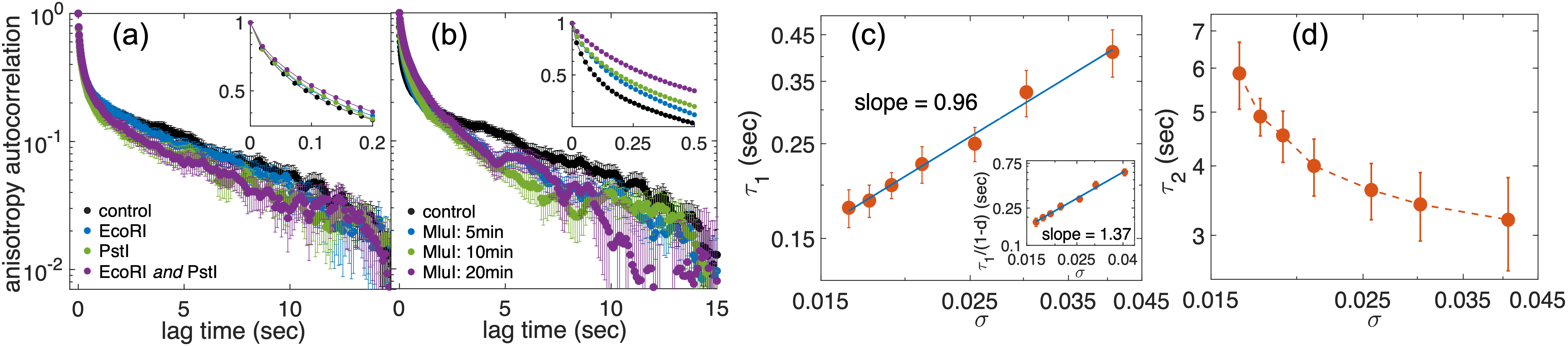}
    \caption{Ensemble averaged time autocorrelation function (a) with well control removal of minicircles/maxicircles, (b) random removal of minicircles/maxicircles from the kDNA network. (c) Master curve ( $\tau_{1}$ versus $\sigma$ ) showing the universal nature of $\tau_{1} \sim \sigma^{0.96}$. (d)  $\tau_{2}$ versus $\sigma$ showing the monotonic decrease of $\tau_{2}$. Inset in (c) presents the influence of drag changing with degree of digestion (d) on scaling exponent. }
    \label{fig:fig4}
\end{figure*}

Figure 4(a) presents the autocorrelation of anisotropy corresponding to the digestion of the minor class of minicircles or/and maxicircles. Inset of the figure shows a zoom in view of the autocorrelation for short lag times. It is clear from the inset that as rings are systematically removed from the network, the time constant for fast relaxation consequently increases. For the slower relaxation, however, the autocorrelation decreases relatively faster as we remove rings from the network, i.e., the time constant for slow relaxation decreases. Similar opposing trends between $\tau_1$ and $\tau_2$ but with more significant differences in magnitude are obtained for molecules after time-dependent digestion of rings using \textit{MluI} (Figure 4(b)).

The first time constant ($\tau_1$) related to the faster decay of the autocorrelation function corresponds to a conformational decorrelation i.e., time over which the network forgets its initial configuration. For a linear polymer, it also corresponds to the timescale over which a molecule diffuses a distance equivalent to its radius of gyration and is the relaxation time for the lowest mode in the Rouse model \cite{Doi1988,Yadav2020}. For free draining, the Rouse model considers the polymer as a set of beads connected by Gaussian springs and predicts the relaxation time as $\tau_{1}=\zeta / k_{t o t}$ where $\zeta$ is the total viscous drag on the chain and $k_{tot}$($=k_{B}T/\Delta x^{2}$)
is the effective spring constant with $\Delta x$ being the extension/contraction in Gaussian springs from its equilibrium size. $k_{B}{T}$ is the thermal energy. Topologically 2D network of kDNA can be modelled as a 2D network of self-avoiding beads connected with Gaussian springs (Figure S4) \cite{Kantor1986,Kantor1987}. Hydrodynamic interactions become decorrelated at length scales proportional to the channel height\cite{jones2013intrachain}, accordingly the Rouse dynamics will dominate the shape fluctuations. Extending the Rouse formalism to 2D network it can be shown that $\tau_{1} \sim \zeta\sigma^{2}/k_{B}T$ \cite{SM2022}. 

The scaling of $\tau_{1}$ with $\sigma$ for molecules digested with two different classes of enzymes is shown in Figure S5. The slopes of the regression line for two sets of samples are very close and they fall onto a master plot with a universal scaling ($\tau_{1} \sim \sigma^{0.96}$) (Figure 4(c)). The nature of this master plot implies that the scaling relation between $\tau_{1}$ and $\sigma$ is independent of the detailed arrangements of the rings and/or perforation withing the catenated network. This observation is in accord with studies of perforated sheets wherein the crumpling critical temperature depends on the perforation area, but is independent of detailed arrangement of perforated holes within the sheet \cite{Yllanes2017}. The relaxation time also encrypts the mechanical strength of the network. It has been shown that bending rigidity $(\kappa)$ and relaxation time ($\tau_{1}$) for vesicles are related as $\kappa=\eta r^{3} / \pi \tau_{1}$, where $\eta$ is solvent viscosity and $r$ is the equilibrium radius \cite{Zhou2011}. 
For a given $\eta$ and $r$, $\kappa \sim \tau_{1}^{-1}$. Hence, removing rings softens the network.  The calculated values of $\kappa$ for different samples are listed in Table S1. 

The deviation from the Rouse prediction ($\tau_{1} \sim \sigma^{2}$) could be due to oversimplistic nature of the spring model of kDNA. While the size of the molecules remains invariant, the number of beads in the modelled network should reduce in proportion to the degree of digestion and hence the effective drag. Influence of digestion on effective drag is discussed in the SM. Inset in Figure 4(c) and Figure S6 show that when we account for this modification in drag, the scaling exponent increases to ($ 1.37-1.52$), which is still smaller than the Rouse prediction.
 It should also be noted that though there are many theoretical and simulation studies available about 2D polymers, \cite{Kantor1987,Helfrich1973,Plischke1988,Bowick2001,Yllanes2017,Polson2021} experimental results are scarce, and the prediction of the statistical models for 2D polymers has not been tested. To the best of our knowledge, this is the first experimental data presenting the universal scaling between two measured quantities for a 2D polymer. 

The time constant ($\tau_2$) associated with second mode of relaxation {\it decreases} as we remove rings from the network, in contrast to $\tau_1$. This  trend could not be rationalized by the spring model of the network in bulk. Our prior work on pristine kDNA conjectured that the longer time constant is the result of microfluidic confinement which supresses  out-of-plane motion and can lead to long-lived local folds along the edge of the molecule\cite{Klotz2020}.  Examples are shown in Figure S7 and S8 \cite{SM2022}. The energy barrier of conformational fold transitions is related to $\kappa$ of the network, which decreases monotonically with fraction of digestion (Table S1). The decline of $\kappa$ will also weaken the sense of confinement for the molecules and facilitate out-of-plane motion. Together, these contributions lead to a reduction of $\tau_2$ upon ring digestion. 



In this Letter, we tuned the topology of a 2D catenated DNA
network to understand its influence on  equilibrium conformations and shape fluctuations. Though the molecular topology has minimal influence on the spatial extension of this 2D network, it significantly affects shape fluctuations. Remarkably, irrespective of details of the molecular topology, the relationship between the time constant of shape fluctuations and variance of shape anisotropy shows a universal scaling. 
Our results provide a route to selectively tune the physical properties of 2D catenated DNA networks and in doing so, we discovered unanticipated trends in properties related to topological modifications which will hopefully spurn further studies of this emergent class of polymers. 



\bibliography{apssamp}

\end{document}